\documentclass[final,authoryear,3p,times,review,10pt]{elsarticle}
\usepackage{graphicx,url}
\usepackage[tight,footnotesize]{subfigure}
\usepackage[font={small,it}]{caption}

\usepackage{amssymb}
\usepackage{amsmath}

\journal{JASIST}

\begin{document}

\begin{frontmatter}



\title{A Multi-dimensional Investigation of the Effects of Publication Retraction on Scholarly Impact}


\author[1]{Xin Shuai\corref{mycorrespondingauthor}}
\cortext[mycorrespondingauthor]{Corresponding author}
\author[1]{Isabelle Moulinier}
\author[2]{Jason Rollins}
\author[1]{Tonya Custis}
\author[1]{Frank Schilder}
\author[3]{Mathilda Edmunds}
\address[1]{Research \& Development Group, Thomson Reuters, 610 Opperman Dr., St Paul, MN 55123}
\address[2]{Intellectual Property \& Science, Thomson Reuters, 50 California St., San Francisco, CA 94111}
\address[3]{Intellectual Property \& Science, Thomson Reuters, 1500 Spring Garden St., Philadelphia, PA 19130}
\address{Email: \{xin.shuai, isabelle.moulinier, jason.rollins, tonya.custis, frank.schilder, mathilda.edmunds\}@thomsonreuters.com}

%


\begin{abstract}
Over the past few decades, the rate of publication retractions has increased dramatically in academia. 
In this study, we investigate retractions from a quantitative perspective, aiming to answer two fundamental questions. One, how do retractions influence the scholarly impact of retracted papers, authors, and institutions? Two, does this influence propagate to the wider academic community through scholarly associations? 
Specifically, we analyzed a set of retracted articles indexed in Thomson Reuters Web of Science (WoS), and ran multiple experiments to compare changes in scholarly impact against a control set of non-retracted articles, authors, and institutions. We further applied the Granger Causality test to investigate whether different scientific topics are dynamically affected by retracted papers occurring within those topics.
 Our results show two key findings: first, the scholarly impact of retracted papers and authors significantly decreases after retraction, and the most severe impact decrease correlates to retractions based on proven purposeful scientific misconduct; second, this retraction penalty does not seem to spread through the broader scholarly social graph, but instead has a limited and localized effect. 
 Our findings may provide useful insights for scholars or science committees to evaluate the scholarly value of papers, authors, or institutions related to retractions. 
\end{abstract}

%
%

\makeatletter
\def\ps@pprintTitle{%
  \let\@oddhead\@empty
  \let\@evenhead\@empty
  \let\@oddfoot\@empty
  \let\@evenfoot\@oddfoot
}
\makeatother

\end{frontmatter}

\section*{Introduction}
Sir Isaac Newton said, ``If I have seen further, it is by standing on the shoulders of giants.'' Scientific progress and communication, mostly presented in the form of scientific publications, have laid a solid foundation for the overall development and advance of science for centuries. Accumulative scientific advances are predicated on the assumption that scholars are honest and serious about the accuracy and integrity of their published work. Unfortunately, this is not always true. The scientific community bears the responsibility of self-examination and self-correction to ensure the integrity and authority of scientific literature.

Over the past few decades, the rate of paper retractions, due to errors or purposeful misconduct, has increased dramatically in nearly all academic fields~\citep{lmichan:VanNoorden2011Science, 10.1371/journal.pone.0068397}. A retraction of a published article indicates that the ideas, methodology, or results presented in the original article are shown to be scientifically invalid, and therefore can no longer serve as the proverbial ``shoulders of giants''.
The most common reasons for retraction are scientific misconduct (i.e. falsification, plagiarism) or unintended errors. Retractions are typically initiated by journal editors or by the article's authors themselves. 

Recently, a growing list of retracted papers has drawn the attention and scrutiny of both the academic and the popular media.
In one widely-reported example, a study about evolving attitudes toward gay marriage published in \emph{Science}~\citep{gaymarriage}, was retracted because one of the authors was not able to provide the raw data. \emph{The Lancet} retracted a study by~\citet{vaccine} that suggested that combined vaccines of measles, mumps, and rubella lead to autism in children. Despite the retraction of the study, many parents continue to believe it, which has resulted in a decline of vaccines for children in Britain and the U.S. In another prominent case, two papers about human stem cells published in \emph{Science}~\citep{stem1, stem2}, were retracted because the authors fabricated the data.


While the systematic and exhaustive study of the publication retraction is still in a nascent stage, several previous studies explore this area from different aspects and inform our work.~\citet{Fang16102012} categorized more than 2,000 biomedical and life science research articles based on retraction reasons and found that more than 60\% of retractions are attributed to misconduct.~\citet{mkarsai:Lu2013Retraction} conducted a controlled experiment showing that the citation counts of retracted articles decrease more rapidly after retraction than those for non-retracted control articles, and prior articles by authors of retracted papers, published before retraction, are also negatively affected.~\citet{ChenHMS13} discussed the use of a visualization tool to study the relation of retracted articles to other scientific literature.

Although these studies exhibit several interesting findings, they are generally limited in scope to narrow facets of the study of retractions. The research presented in this paper is the first attempt at a comprehensive and systematic investigation of the phenomenon of retraction. More ambitiously, we attempt to frame the phenomenon of retractions in the broader context of the overall impact on academia and the scientific community. To be specific, 
\begin{itemize}
\item we not only measure the effect of paper retraction on articles and scholars, but also the effect on institutions; 
\item we not only explore the impact change of retracted articles and scholars, but also the impact change of articles and scholars that are directly or indirectly related to retracted articles and scholars;
\item  we not only investigate specific retraction cases within certain research topics, but also the potential effect of retractions on the future popularity of those topics.
\end{itemize}
 Overall, we conduct our research from two perspectives: One, how do retractions affect the scholarly impact of papers, authors, and institutions? Two, does such an effect propagate to the wider scientific or academic community through scholarly association? The main contributions of this paper include:

\begin{itemize}
\item The design of a coding schema to categorize retracted papers based on retraction reason. 
\item The design of multiple statistical comparative experiments to study the effect of retractions on the scholarly impact of articles, scholars, and institutions.
\item The investigation of temporal causal correlation between paper retraction rate and publishing popularity within certain scientific topics. 
\end{itemize}


\section*{Related Work}
The phenomenon of publication retraction in academia has recently increased. Although the scientific investigation of retraction is still at an early stage, several studies attempt to explore the characteristics of retracted papers and their effect on scholarly impact.

Even though the increasing number of retraction cases have received attention from several scholars~\citep{lmichan:VanNoorden2011Science, 10.1371/journal.pone.0068397}, the scientific community has yet to pay full attention to it~\citep{lmichan:VanNoorden2011Science, doi:10.6087/kcse.34}, and retracted papers are still being cited~\citep{Pfeifer}.~\citet{Fang16102012} categorized more than 2,000 biomedical and life science research articles based on retraction reason and found that more than 60\% of retractions were attributed to misconduct.~\citet{network} proposed a more complete categorization schema and found out that most citations of retracted papers after retraction are from the authors themselves. In addition to categorization,~\citet{ChenHMS13} discussed the use of a visualization tool to study relationships between retracted articles and other scientific literature.
In addition, there are several studies that attempt to find factors that correlate with retractions.~\citet{Repec} suggest that attention is a key predictor of retraction, and that retractions arise most frequently among highly-cited articles.~\citet{10.1371/journal.pone.0127556} point out that the key factors leading to scientific misconduct are lack of research integrity policies, money-driven publishing behavior, and career pressure for junior researchers.

Many studies describe and quantify the aftermath of paper retractions. Some researchers focus on specific retraction cases.~\citet{Bornemann-Cimenti:2015aa} examine the citation change of scholar Scott Reuben's articles within 5 years of their retraction, and found that most of this retracted work is still being cited without mention of retraction by other scholars.~\citet{10.1371/journal.pmed.1000318} defined a mathematical equation to measure the potential economic cost of scientific misconduct by a federal grant applicant.

Other researchers extended the investigation from specific cases to a set of retracted articles or authors.
~\citet{mkarsai:Lu2013Retraction} conducted a controlled experiment showing that the citation counts of retracted articles decrease more rapidly after retraction than for non-retracted control articles. Prior articles by authors of retracted papers, published before retraction, are also negatively affected.~\citet{Pierre1, NBERw21146} applied a similar controlled experiment methodology and showed evidence for the negative effects of retractions on papers and their authors? citation count. They also showed that retractions due to fraud lead to a more severe negative effect than honest errors.~\citet{ASI:ASI23421} further confirmed the negative impacts to scholars involved in a retracted article.

Our work aligns with and extends these lines of research by examining the effects of retraction on the scholarly impact of articles, scholars and institutions from a more systematic and comprehensive perspective. We further investigate whether such effect propagates to the wider scientific or academic community through scholarly associations.

\section*{Data Description and Overview}
Our dataset comprises records from Thomson Reuters Web of Science (WoS) based on articles published from 1980 to 2014 across a wide range of research disciplines. For each article, metadata including Title, Authors, Published Date, Journal, ESI category\footnote{ESI category is a journal subject categorization system coined by Thomson Reuters,~see~\url{http://ipscience-help.thomsonreuters.com/incitesLive/7622-TRS/version/1/part/5/data/ESI_Journal_Category_Map_2012.xlsx?branch=incites_115&language=en_US}}, Institutions, and Cited References were used in our analysis. To identify retracted publications, we searched for titles containing ``Retracted article'' (e.g.~Arabidopsis Downy Mildew Resistance Gene RPP27 Encodes a Receptor-like Protein Similar to CLAVATA2 and Tomato Cf-9 (Retracted Article. See vo.l 143, pg. 1079, 2007)~\cite{Arab})).  

\begin{figure}[!ht]
 \centering
 	\subfigure[]{
	    \label{fig:retraction_rate}
		\includegraphics[width=0.317\textwidth]{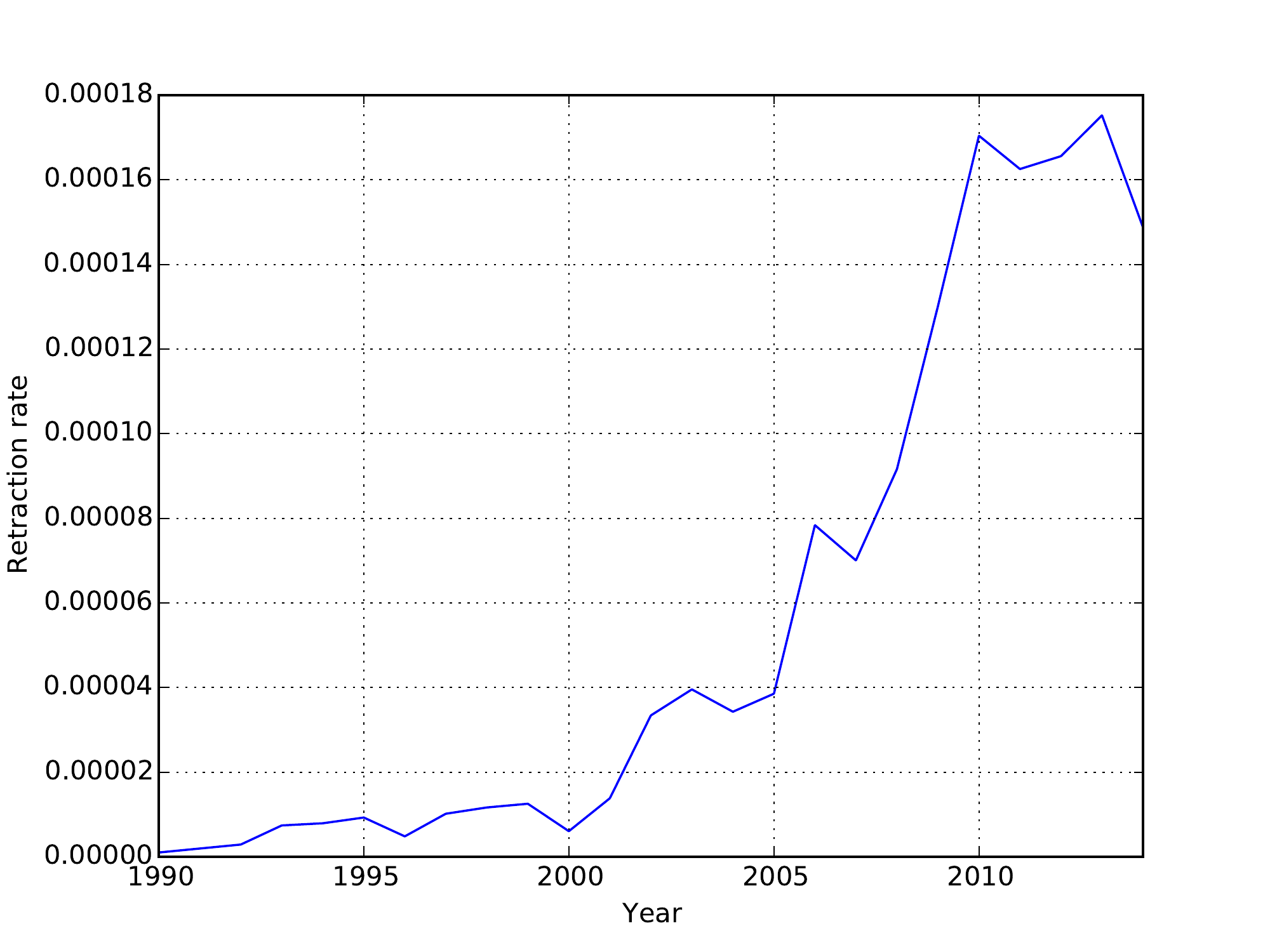}
	}
	\subfigure[]{
	   \label{fig:retraction_cit}
	   	\includegraphics[width=0.317\textwidth]{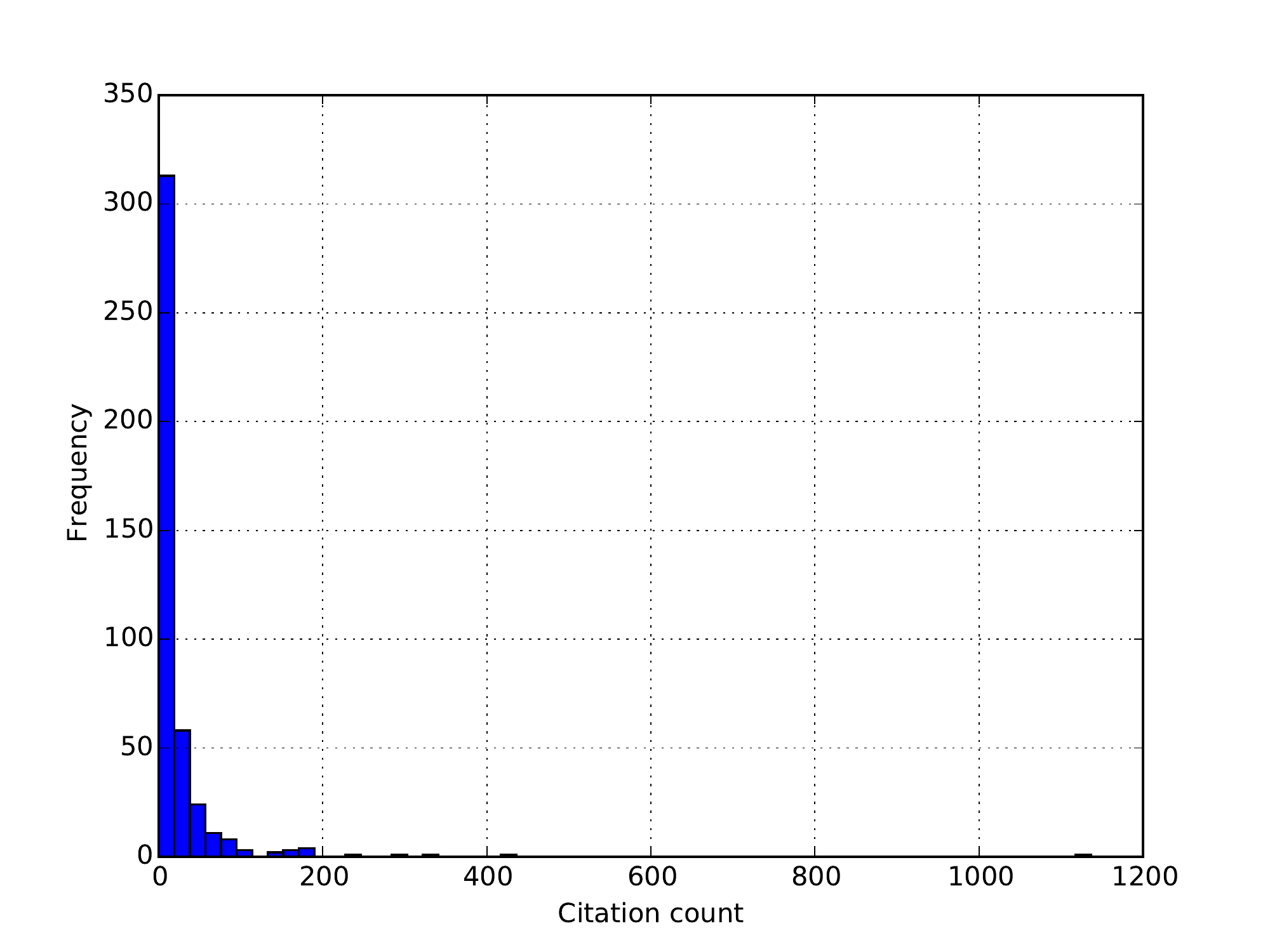}
	}
	\subfigure[]{
	   \label{fig:retraction_dist}
	   	\includegraphics[width=0.317\textwidth]{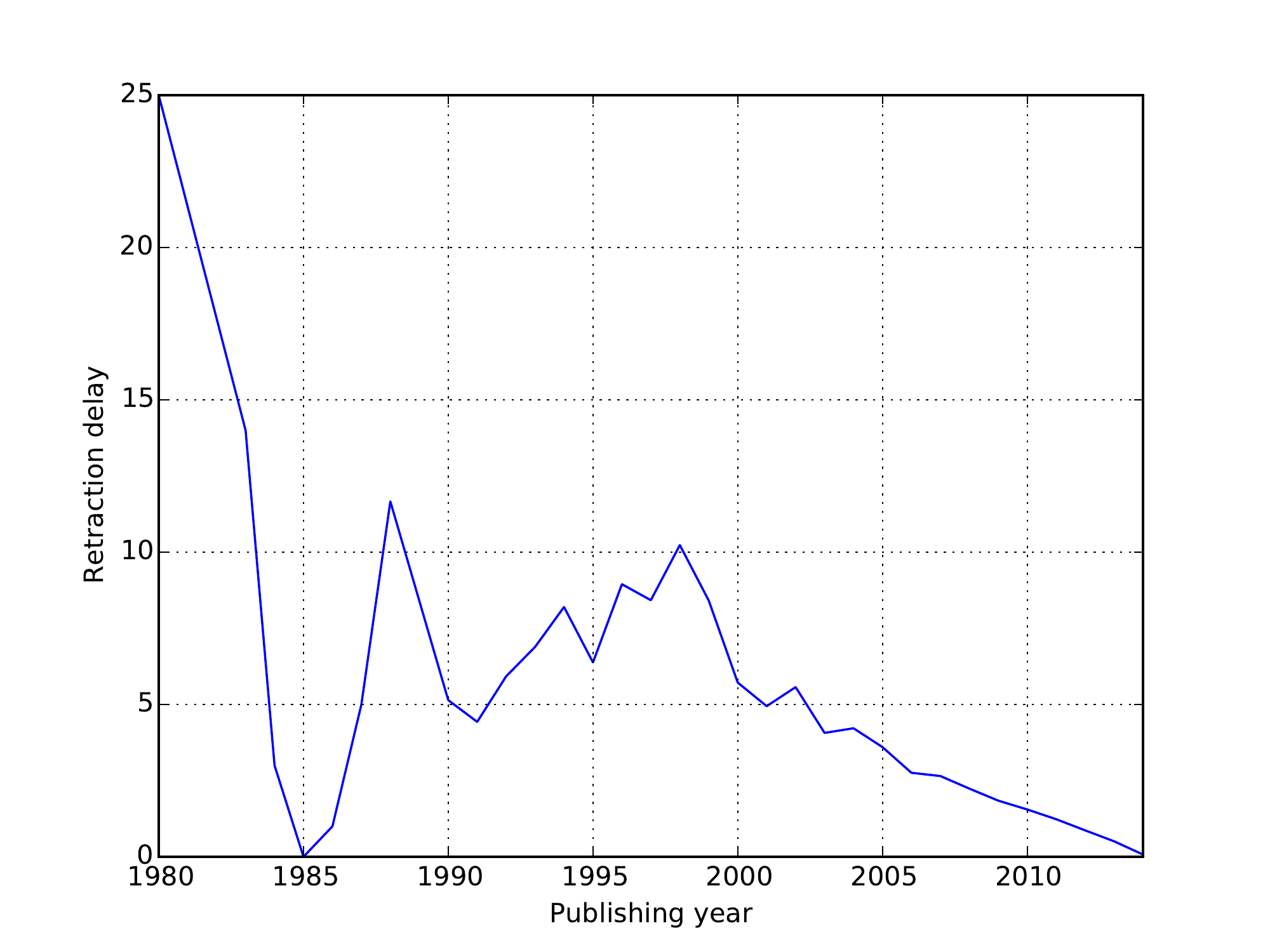}
	}	
	
\caption{Several statistics about retracted papers. Specifically, (a) shows the dynamic change of retraction rate, (b) shows the distribution of citation counts of retracted papers and (c) shows the dynamic change of retraction delay}\label{fig:retracted} 
\end{figure}

In total, we extracted 2,659 retracted articles. The change of annual retraction rate, which we define as the ratio of the number of retracted articles to the total number of published articles in a certain year, is shown in Figure~\ref{fig:retraction_rate}. 
We can see that the annual retraction rate has increased by almost 20 times over the past 25 years.
Table~\ref{tab:top_ret} shows the top five, as well as bottom five scientific ESI categories ranked by retraction ratio (i.e., the number of papers retracted over the total number of papers published in that category).
\begin{table}
  \centering
  \scalebox{0.9}{
             \begin{tabular}{c|c|c} \hline
                Rank  						&ESI Category					&Retraction Rate ($\times 10^{-4}$)\\
                \hline\hline
                1							&molecular biology \& genetics		&2.1\\
                2							&immunology					&1.66\\
                3							&microbiology					&1.55\\
                4							&biology \& biochemistry			&1.14\\
                5							&pharmacology \& toxicology       	&1.08\\
                \hline\hline
                18							&plant \& animal science			&0.36\\
                19							&social sciences, general			&0.21\\
                20							&geosciences					&0.17\\
                21							&computer science				&0.17\\
                22							&space science					&0.06\\
                \hline
             \end{tabular}}\centering
      		\caption{Top five and bottom five frequently retracted research fields defined in ESI categories}
        \label{tab:top_ret}
\end{table}
Notably, high retraction rates mostly occur in medical or biological related research fields.
Figure~\ref{fig:retraction_cit} shows the distribution of the citation counts of retracted articles and the distribution is very skewed. We also found that the median citation count of retracted articles is eight, while the median count for all articles is only one, indicating that a retracted article is generally cited more than an average article. 



In addition, we looked at retraction delay, which we define as the difference between the publishing year and the retraction year. The median retraction delay among all retracted articles is two years. Figure~\ref{fig:retraction_dist} shows that the retraction delay has greatly shortened in recent years, indicating that the scientific community is responding to retract studies faster than ever before. We hypothesize that the increased speed of retractions may be due to the development of digital libraries and online publishing that facilitate and accelerate scholarly communication. Consequently, errors in scientific literature, whatever intentional or not, can be identified by a wider scientific community much easier and faster.


\section*{Concepts and Definitions}

This section is organized as follows: First, we design a coding schema to categorize the reasons for retraction, which enables us to investigate how different types of retraction lead to different degrees of effect on scholarly impact. Second, we define retracted papers, authors and institutions, as well as related papers and authors, in terms of citation relations. 
Finally, we describe the concepts of research topics, topical popularity and retraction rate, in order to examine possible temporal correlations between retractions and topic popularity. 
\begin{figure}[htbp]\centering
	\includegraphics[width=0.85\columnwidth]{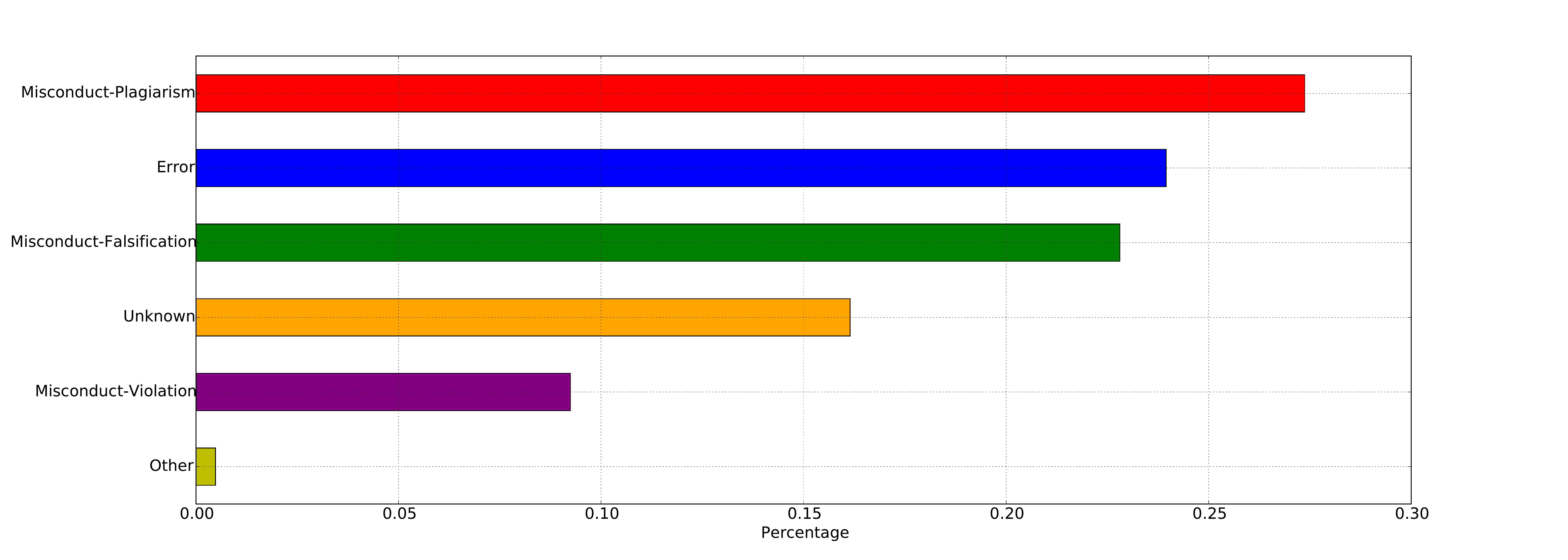}
                \caption{The distribution of reasons for paper retraction. The most frequent reason for retraction is \emph{plagiarism}, followed by \emph{error} and \emph{falsification}.}
                \label{fig:reason}
\end{figure}

\subsection*{\textit{Coding schema for retraction categorization}}
Scholarly publications are retracted for several reasons. Typically, either the authors of the paper or the editor of the publishing journal can request a retraction. Understanding the reason for retraction is useful in scholarly impact analysis; we hypothesize that different types of retraction have different degrees of effect on post-retraction impact. After a preliminary investigation of a sample set of retracted papers, we devised a classification schema adapted from the Wikipedia page for Scientific misconduct\footnote{\url{https://en.wikipedia.org/wiki/Scientific_misconduct}}:
\begin{itemize}
\item Scientific misconduct: the intentional violation of standard codes of scholarly conduct and ethical behavior in scholarly publication of scientific research. It can be further divided into three sub-categories:
		\begin{enumerate}
		\item Plagiarism: copying the ideas, contents (including text, figures, tables, data), and results of others without explicit citation; duplicate publication of the same materials in different journals.
		\item Falsification or Fabrication: making up results or data, manipulating research materials, process, or data so that the research is not accurately represented.
		\item Violation of rules: submitting articles without approval of all co-authors; conducting human experiments without IRB approval; or other types of violations not directly related to the content of articles. 
		\end{enumerate}
\item Errors: unintentional errors made by authors during the process of data collection, processing, or analysis causing the final results to be invalid. Errors can also occur during the process of article publishing. 
\item Others: any other reasons that are not due to the fault of the authors. For instance, the publishers may unintentionally publish two versions of the same article and need to retract one version.  
\item Not found: the retraction notice was not found, or the reason for retraction was not explicitly given.
\end{itemize}
In addition, we categorized three types of retraction requests: editor's request, author's request, or not found.  

Four raters, including three authors of this study, annotated 1,666 of 2,659 retracted articles, tagging who requested the retraction and the reason for retraction. To quantify the inter-rater agreement, a set of 100 articles are randomly selected and assigned for annotation to all four raters. The calculated Fleiss's kappa is 0.73, indicating that the degree of inter-rater agreement is high and the annotation results are trustworthy.  
Figure~\ref{fig:reason} shows the distribution of reasons for retraction. More than 25\% of retraction cases are due to plagiarism (including duplicate publications). The second-most prevalent reason for retractions (around  24\%) is unintentional author errors in the process of experimentation or data analysis. The third-most prevalent reason for retractions (around 23\%) is falsification and fabrication. Both plagiarism and falsification are considered egregious forms of scientific misconduct. 

\begin{figure}[htbp]\centering
	\includegraphics[width=0.7\columnwidth]{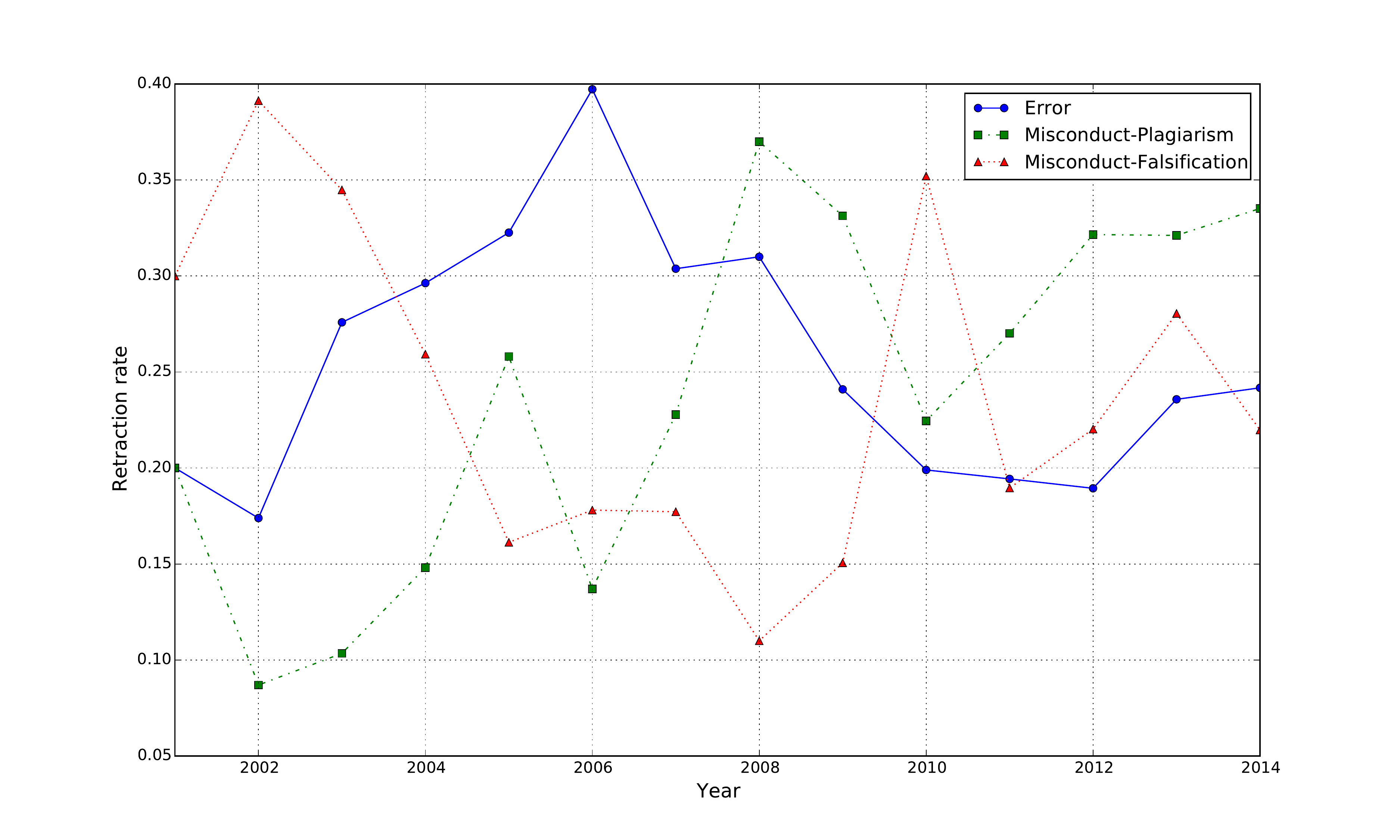}
                \caption{Yearly dynamic change for the top three retraction reasons. The retraction rate due to \emph{Plagiarism} seems to be increasing, while the trends for \emph{Error} and \emph{Falsification} show one or two spikes, but no overall increase.}
                \label{fig:dynamic}
\end{figure}

Figure~\ref{fig:dynamic} shows the dynamic yearly change, starting in 2000, of the top three retraction reasons shown in Figure~\ref{fig:reason}, \emph{plagiarism}, \emph{error}, and \emph{falsification}. According to our observation, the retraction rate due to \emph{plagiarism} is increasing, while the trends for retraction rates due to the other two reasons, show one or two spikes (\emph{error} in 2006, \emph{falsification} in 2002 and 2010). We hypothesize that \emph{plagiarism} has become the most frequent type of misconduct due to the availability of much more online scientific literature to ``borrow from'' than ever before.

\subsection*{\textit{Retracted entities and related entities}}
First, we define the main concepts as follows:
\begin{itemize}
\item \textbf{Academic entity}: Paper, author or institution in the WoS corpus. Each paper is assigned a unique ID, while authors and institutions are each represented by a unique name string after disambiguation and normalization. 
\item \textbf{Scholarly impact}: Total number of citation counts. Scholarly papers are connected through citations, and the citation count is one of the most recognized indicators of scholarly impact. Similarly, the scholarly impact of authors and institutions can be represented by the sum of citation counts of their corresponding papers.
\item \textbf{Impact curve}: The temporal curve of yearly citation counts. If we use $C(y)$ to denote the scholarly impact of any entity in year $y$, then the \emph{impact curve} can be formulated as a temporal curve within specific period: $C(y), y_0\leq y \leq y_n$. We set $y_n$ to 2014, while $y_0$ has different definitions based on the type of entity. For each paper, $y_0$ is the publication year; for each author or institution, $y_0$ is the first year when the author or institution is cited, i.e. the year that the scholarly impact begins to build up.
\item \textbf{Retraction year}: The first year when an academic entity was involved in a retraction. For different entities, the specifications of retraction year are different, and the details will be shown later.
\item \textbf{Retracted entity}: Academic entity that was involved in a retraction case. Specifically, once a paper was announced as retracted, all authors of this paper, as well as the institutions those authors were affiliated to, are considered retracted authors and retracted institutions.
\end{itemize}
Scholarly impact is not a static measure, but dynamically changing all the time as new publications emerge. The main question under investigation in this paper is how formal retractions affect the scholarly impact of involved and related papers, authors and institutions.

\subsubsection*{\textit{Scholar and institution name disambiguation}}
Each article extracted from the Thomson Reuters Web of Science (WoS) collection has a unique ID field. In contrast, authors and institutions are represented only by name strings, which often need to be disambiguated.

We use two proprietary Thomson Reuters tools for disambiguation. The first tool~\citep{griffith2015method} uses a semi-supervised machine learning algorithm that clusters all author names in WoS into different groups with each group corresponding to one unique scholar; this reaches 95\% precision and 84\% recall. For institution name disambiguation, we used the Web Application for Address Normalization~\citep{waan}, which normalizes to the root-organization level (e.g. University), but not to the sub-organization level (e.g. Department or School). 

\subsubsection*{\textit{One retraction example}}
\label{sec:example}
\begin{figure}[h!]
 \centering
 	\subfigure[]{
	    \label{fig:doc_curve}
		\includegraphics[width=0.319\textwidth]{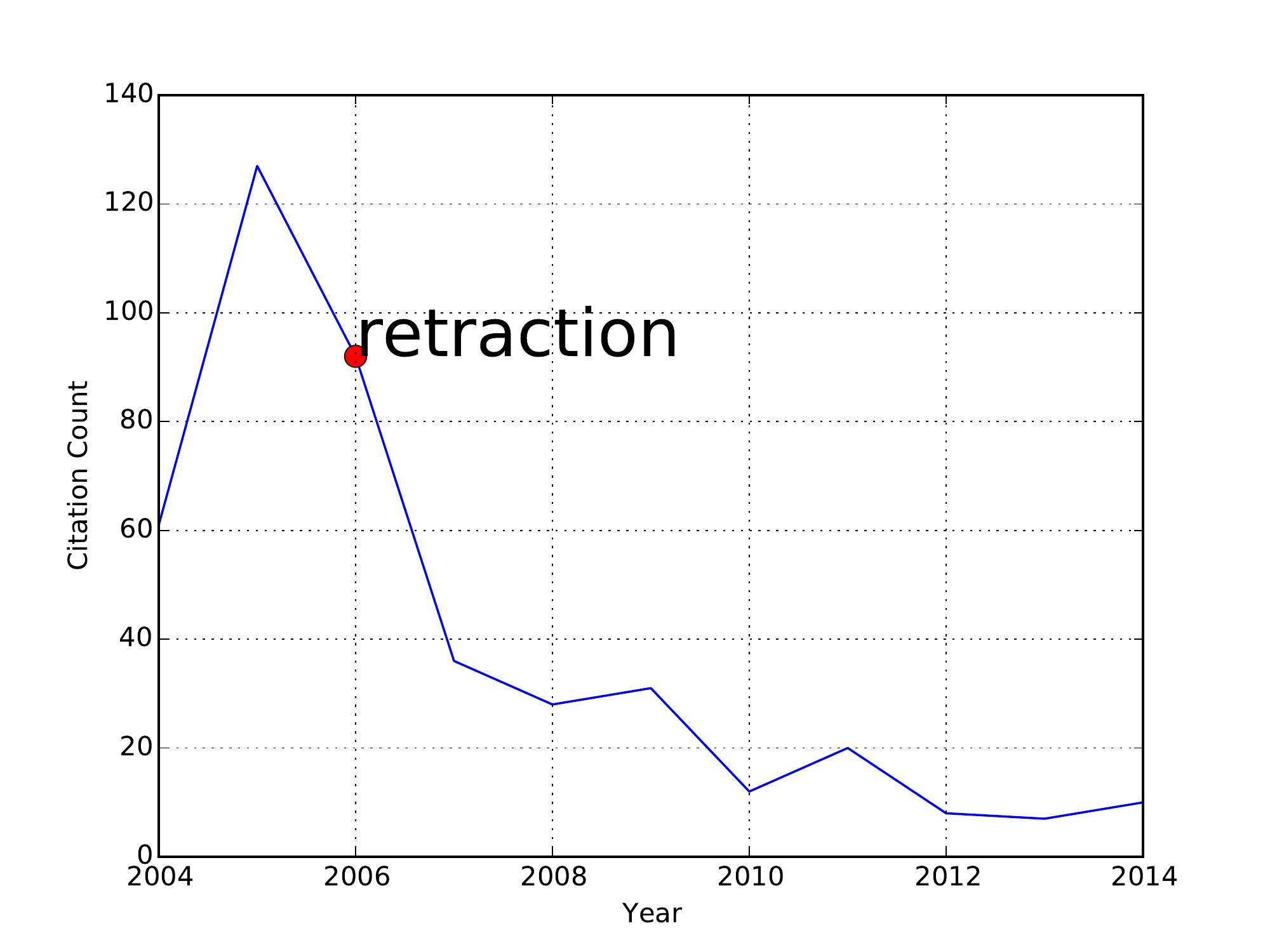}
	}
	\subfigure[]{
	   \label{fig:author_curve}
	   	\includegraphics[width=0.319\textwidth]{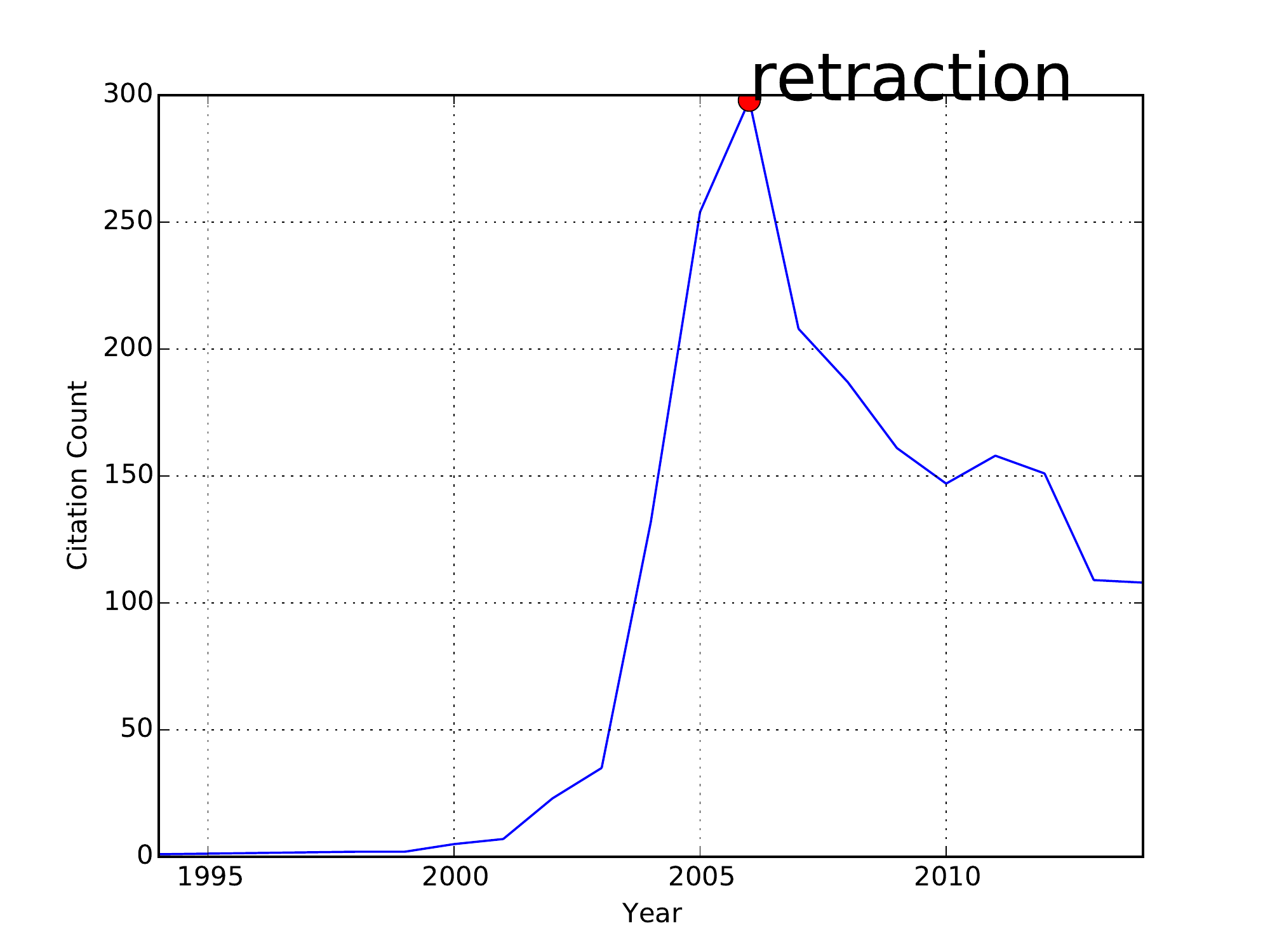}
	}
	\subfigure[]{
	   \label{fig:inst_curve}
	   	\includegraphics[width=0.319\textwidth]{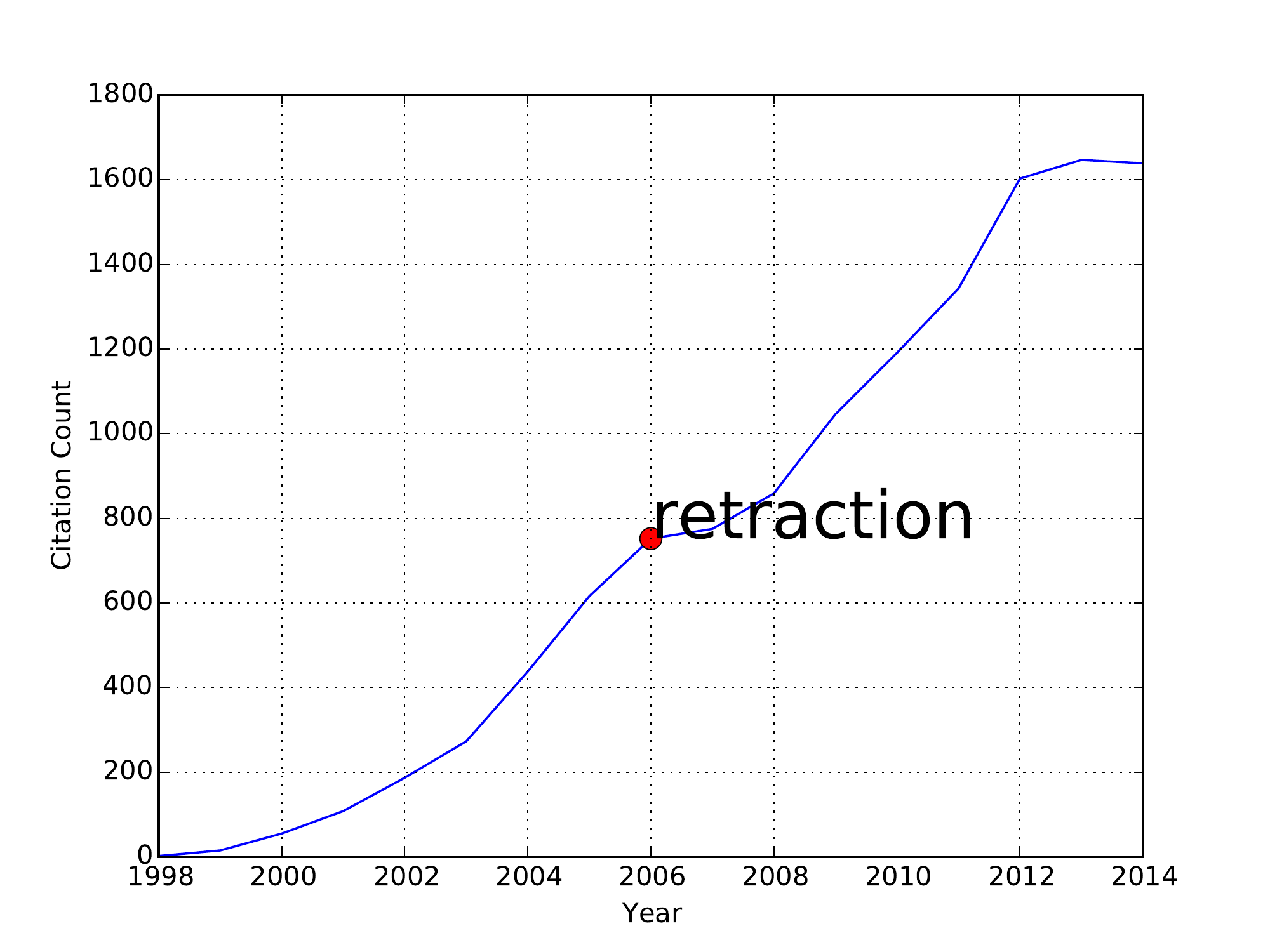}
	}	
	

\caption{The impact curves of (a) the 2004 stem cells paper (b) \emph{Hwang, Woo Suk} and (c) \emph{Seoul National University College of Veterinary Medicine}. In (a) and (b) we can spot the decreasing trend after the retraction point.}\label{fig:example_curve} 
\end{figure}

Two papers about cloning and human stem cells were published in \emph{Science}~\citep{stem1, stem2} and retracted in 2006, because much of the data were found to be fabricated. The name of both papers' first author is \emph{Hwang, Woo Suk}, who was a professor at \emph{Seoul National University College of Veterinary Medicine}. Figure~\ref{fig:example_curve} shows the impact curves for the retracted paper published in 2004, the author, and the institution. We can clearly spot the decrease of scholarly impact after retraction for the retracted paper and the author (See Figure~\ref{fig:doc_curve} and \ref{fig:author_curve}). However, such a decrease is not shown for the institution (See Figure~\ref{fig:inst_curve}). We conduct statistical comparative experiments to test the significance of these decreases in citation count in Section 5 of this paper.

\subsubsection*{\textit{Related entities to retraction}}
In addition to the explicitly retracted entities, effects on other academic entities that are related to the retracted entities are also within the scope of our investigation. For retracted papers, other papers that cite the retracted papers (i.e. citing relation), or share the same references as the retracted papers (i.e. co-reference relation) are considered to be related papers. A related author (i.e. co-authorship relation) is defined as an author who once co-authored with a retracted author, but on a non-retracted paper. We do not consider related institutions here, since it is not easy to define and measure the relatedness between institutions. By exploring changes in scholarly impact of related, but non-retracted, papers and authors, we attempt to address whether the retraction effect spreads through the scientific community through scholarly association.
Table~\ref{tab:entity} summarizes the six types of retracted and related entities that will be examined.
\begin{table}
  \centering
  \scalebox{0.9}{
             \begin{tabular}{l|l} \hline
                Entity 													&Variable/Symbol\\
                \hline\hline
                Retracted papers										&$P$\\\hline
                Retracted authors											&$A$\\\hline
                Retracted institutions											&$I$\\\hline
                Papers citing retracted papers							&$P_{citing}$\\\hline
                Papers sharing references with retracted papers				&$P_{coref}$\\\hline
                Authors co-authoring with retracted authors in non-retracted papers	&$A_{coaut}$\\\hline
             \end{tabular}}\centering
      	\caption{Six types of entities that will be investigated on their scholarly impact change after retraction.}
        \label{tab:entity}
\end{table}

\subsection*{\textit{Research topics and retraction}}
\begin{figure}[h!]
 \centering
 	\subfigure[]{
	    \label{fig:gene}
		\includegraphics[width=0.48\textwidth]{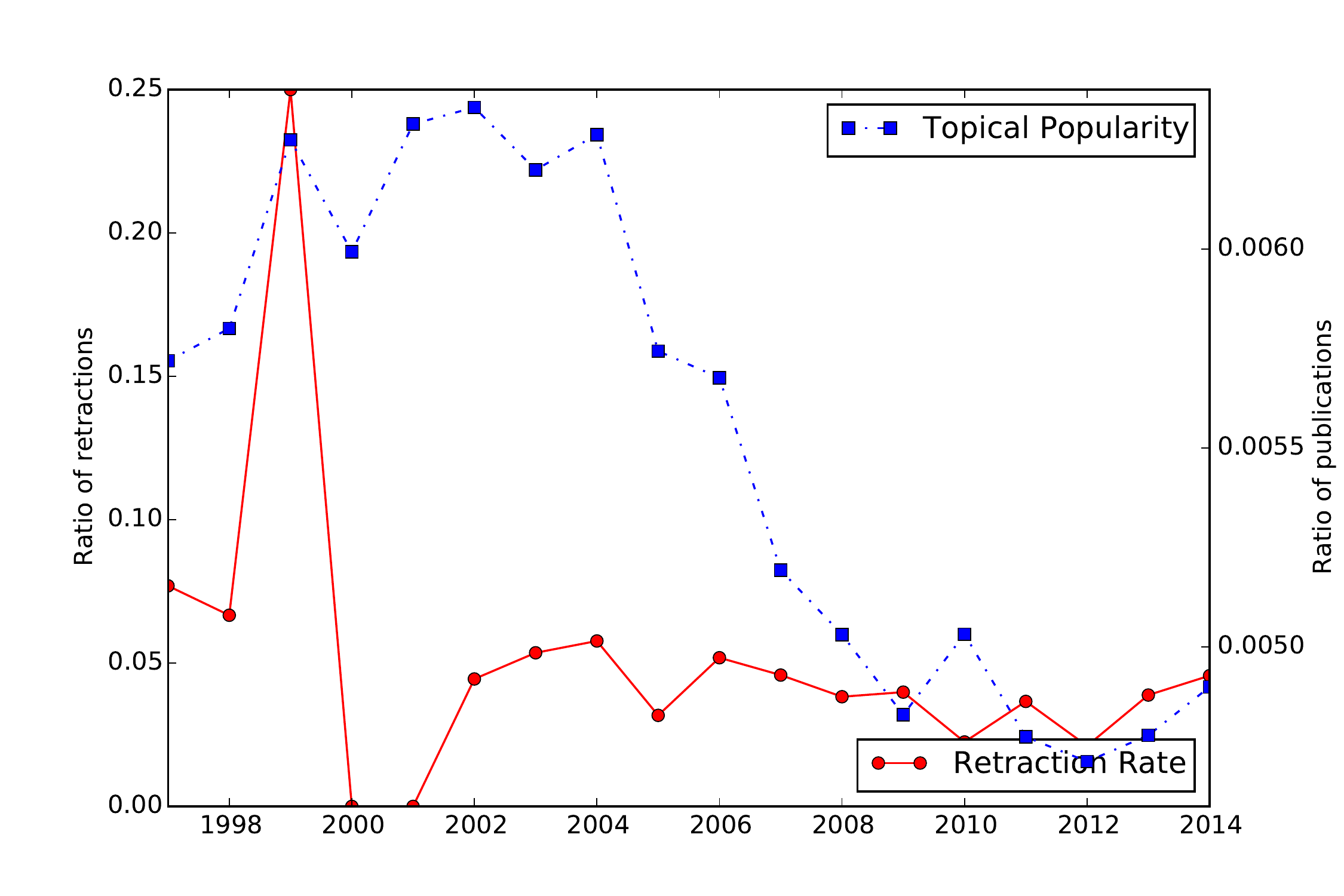}
	}
	\subfigure[]{
	   \label{fig:apop}
	   	\includegraphics[width=0.48\textwidth]{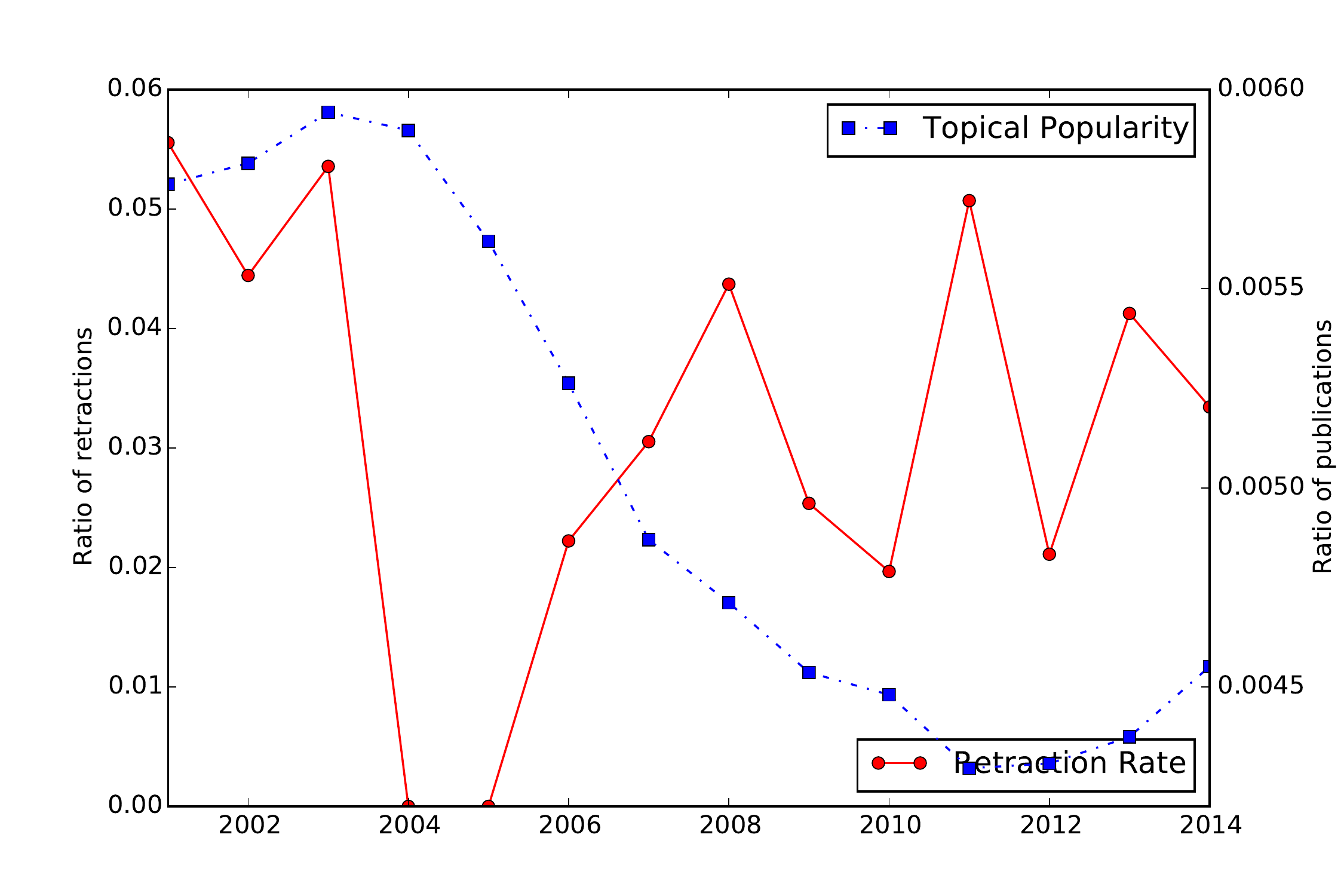}
	}

\caption{The trend of topical popularity ($Pop^k(y)$) and retraction rate ($Ret^k(y)$) of two topics.}\label{fig:topic} 
\end{figure}

The academic entities illustrated above in Table~\ref{tab:entity} can also be grouped by scientific topics. Here, the scientific topic refers to a specific research direction within a more general scientific discipline (defined by ESI category). For instance, \emph{Stem Cell} is a hot topic within \emph{Molecular Biology}, and \emph{Big Data} is an emerging topic in \emph{Computer Science}. Some key concepts are listed as follows:
\begin{itemize}
\item \textbf{Scientific topic}: Given a paper, we assign scientific topics based on the title of that paper; we use a short text annotation tool \emph{Tagme}\footnote{\url{http://tagme.di.unipi.it/}} to detect Wikipedia concepts from all titles of retracted papers. One Wikipedia concept corresponds to one Wikipedia page, and the title of the page is the name of the detected topic. We assume that all extracted topics constitute a topic set $K$.
\item \textbf{Yearly publication count}: Given a year $y$, the total number of papers found in our WoS collection in this year is denoted as $Pub(y)$.
\item \textbf{Yearly topical popularity}: Given a year $y$ and a topic $k\in K$, the total number of papers that belong to topic $k$ is denoted as $Pub^k(y)$. Then the topical popularity of topic $k$, can be measured as $Pop^k(y)=Pub^k(y)/Pub(y)$
\item \textbf{Yearly topical retraction rate}: Given a year $y$, the total number of retracted papers that belong to topic $k$ is denoted as $Pub^{rk}(y)$. Then we can define a yearly topical retraction rate of topic $k$ as $Ret^k(y)=Pub^{rk}(y)/Pub(y)$
\end{itemize}
From the titles of the 2,659 retracted articles we studied, we extracted more than 4,000 Wikipedia concepts (i.e. scientific topics). For each extracted topic, we assign its ESI category as the most frequently appeared ESI category from all of its associated papers.
The top ten concepts in terms of frequency are listed in Table~\ref{tab:top}.
\begin{table}
  \centering
  \scalebox{0.9}{
             \begin{tabular}{c|c|c} \hline
                Topic keyword  					&ESI category				&Frequency\\
                \hline\hline
                regulation of gene expression 		&biology \& biochemistry		&107\\
                gene expression				&biology \& biochemistry		&92\\
                protein						&biology \& biochemistry		&91\\
                apoptosis						&clinical medicine			&88\\
                cell (biology)					&clinical medicine			&77\\
                gene							&molecular biology \& genetics	&68\\
                enzyme inhibitor				&clinical medicine			&52\\
                cancer						&clinical medicine			&48\\
                tumor							&clinical medicine			&41\\
                inflammation					&clinical medicine			&38\\
                \hline
             \end{tabular}}\centering
      		\caption{Top ten most frequent scientific topics from retracted papers.}
        \label{tab:top}
\end{table}
For each topical keyword $k$, we construct two time series, the yearly retraction rate $Ret^k(y)$ and the yearly topical popularity $Pop^k(y)$. Figure~\ref{fig:topic} shows $Ret^k(y)$ and  $Pop^k(y)$ for the top two frequently occurring topics belonging to different ESI categories: \emph{apoptosis} and \emph{regulation of gene expression}. Although there's no obvious temporal correlation or pattern between $Pop^k(y)$ and $Ret^k(y)$ shown in Figure~\ref{fig:topic}, we examine their possible temporal correlation through statistical significance testing in Section 5 of this paper.

\section*{Experimental Design and Results}
In this section, we conduct a series of statistical experiments in order to examine and quantify the effects of retraction from various aspects. First, we design comparative experiments to test the effects of retraction on three types of academic entities, i.e. papers, authors, and institutions. Second, we study the Granger-causality correlation between the occurrences of retraction within a scientific topic, and the future popularity of that topic.  
\subsection*{\textit{Statistical comparative experiments for retracted or related entities}}
Statistical comparative experiments are used here to test the effect of retraction on the entities listed in Table~\ref{tab:entity}: $P,A,I,P_{citing},P_{coref},A_{coaut}$. Although the implementation details are different for each of the six entities in Table~\ref{tab:entity}, the basic steps used to design and conduct these comparative experiments are:
\begin{enumerate}
\item Select treatment entity type from Table~\ref{tab:entity}.
\item Select treatment entities from pool of retracted entities.
\item Specify the retraction year.
\item Select control entities based on impact curve before the retraction year.
\item Run statistical test to determine if the difference between treatment and control entities is statistically significant.
\end{enumerate}

\subsubsection*{\textit{Selecting treatment entities and specifying retraction year}}
Table~\ref{tab:treatment} shows the criteria for selecting the treatment entities group $E^t$ and the retraction year for each type of entity.
Although the methods of selecting treatment groups differ, we pick entities that are most likely to be affected by the publication retraction intuitively. For instance, the first author of a retracted paper is presumably more affected than the other authors. The paper sharing the most references with a retracted paper is an indicator of the highest topical similarity, thus more likely to be affected as well. The most frequent co-authorship relation with a retracted author is a sign of the closest scholarly relationship with that author, thus more likely to be questioned by academic peers.

Using Table~\ref{tab:treatment} and the example in Section~\ref{sec:example}, we illustrate how treatment entities are selected. The retracted paper~\citep{stem1} itself is assigned to $P^t$. The first author \emph{Hwang, Woo Suk} and his corresponding institution \emph{Seoul National University College of Veterinary Medicine} are assigned to $A^t$ and $I^t$, respectively. From all articles that cited the retracted paper, the first one published~\citep{Eridani:2004aa} is assigned to $P_{citing}^t$. From all articles that share references with the retracted paper, we select the one with the largest Jaccard coefficient between both reference sets~\citep{Tabar:2002aa} and assign it to $P_{coref}^t$. Finally, we find a researcher named \emph{Kang, Sung Keun} who co-authored most frequently with \emph{Hwang, Woo Suk} and assign him to $A_{coaut}^t$. The difference between the two scholars is that \emph{Kang, Sung Keun} did not publish any retracted works.

\begin{table}
  \centering
  \scalebox{0.80}{
             \begin{tabular}{l|l|l} \hline
                Treatment group 			&Selection criterion									&Retraction year\\
                \hline\hline
                $P^t$						&$P^t=P$											&The year of announcement of retraction notice\\
                $A^t$						&The \emph{first author} from $A$					&The retraction year of the author's first retracted paper\\
                $I^t$						&The institution associated with $A^t$					& The retraction year of the institution's first retracted paper\\
                $P_{citing}^t$				&The earliest paper that cites $P^t$					& The retraction year of the cited retracted paper\\
                $P_{coref}^t$				&The paper that shares the most references with $P^t$ 	& The retraction year of the co-referring retracted paper\\
                $A_{coaut}^t$				&The most frequently collaborating co-author with $A^r$		& The retraction year of the collaborating retracted author\\
                \hline
             \end{tabular}}\centering
      	\caption{The method of selecting treatment group and determining retraction year.}
        \label{tab:treatment}
\end{table}

\subsubsection*{\textit{Selecting control entities}}
The basic principle of selecting control entities $E^c$ is to ensure the highest degree of similarity between $e^c\in E^c$ and the treatment counterpart $e^t\in E^t$ previous to the retraction year. In so doing, we are able to more confidently attribute the post-retraction differences to the occurrence of retraction.
Based on the impact curve $C(y)$, we further define a pre-retraction distance $PreDis$ between a treatment entity $e^t$ and a control entity $e^c$ as follows:
\begin{equation}
PreDis(e^t, e^c) = \left \|  C^{e^t}(y) - C^{e^c}(y)\right \|_2, y_0\leq y\leq y_r,
\end{equation}
where $y_r$ is the retraction year. Additionally, we require that $e^t$ and $e^c$ share the same starting year $y_0$. 

In addition to the impact curve and the pre-retraction distance, there exist some other metadata associated with each publication that can help us to select control entities, such as publication date, journal, and ESI category. Authors and institutions do not have ESI category information, but we can assign each author and institution the most frequently appeared ESI categories from all of their publications.
Given a treatment entity $e^t$, we adapt the method from Lu et.~al~\cite{mkarsai:Lu2013Retraction} to select control entities as follows:
\begin{enumerate}
\item Pre-select a set of candidate control entities. Specifically, we ensure that the candidate control papers share the same publication date and publication journal as $e^t$; and the candidate control authors and institutions share the same ESI category and starting year $y_0$ of the impact curve of $e^t$.
\item For all entities in the candidate control set, we pick the top ten entities with the minimum pre-retraction distance ($PreDis$) from $e^t$.
\item From the above selected top ten minimum-scoring entities, we further pick the two entities, $e^{o1}$ and $e^{o2}$, whose citation counts before retraction are the closest to $e^t$, as the final selected control entities for $e^t$
\end{enumerate}

\subsubsection*{\textit{Significance testing}}
Once the treatment and control groups are determined, we use two metrics to compare them: \emph{post-retraction impact} and \emph{impact change ratio}. Pre-retraction, or Post-retraction impact is defined as the sum of citation counts, before or after, the retraction year; impact change ratio is defined as the ratio of post-retraction impact to pre-retraction impact. We test whether the differences for each of the two metrics between the groups are statistically significant. If so, we claim that retraction does affect the scholarly impact of the involved academic entities. During the process of control entities selection, each treatment entity $e^t$ is associated with two control entities $e^{o1}$ and $e^{o2}$. Therefore, we compute the mean post-retraction impact, or mean impact change ratio, of $e^{o1}$ and $e^{o2}$, and compare it with the corresponding single value of $e^t$.

Rather than the more commonly used \emph{t-test} to compare group means between normally distributed samples, we choose to use the \emph{Mann-Whiteney U test} (i.e. nonparametric version of \emph{t-test}) to compare the median values of the treatment and control groups, since the distribution of citation counts is highly skewed (see Figure~\ref{fig:retraction_cit}).

\subsubsection*{\textit{Comparative test results}}
Table~\ref{tab:compare} lists the comparative experimental results for retracted and related entities, and their corresponding selected control entities. The results show that the impact of retracted publications and authors significantly decreased after retraction, indicating that the the scientific community gradually withdraws academic recognition to retracted papers and authors. Interestingly, the scholarly impact of institutions involved in retractions is consistently higher than that of their control group. We hypothesize that this is because many retracted institutions are well-established institutions with good reputations, like Harvard University and MIT, and that retractions from a specific professor or research group hardly affect the reputation of the whole department or university. Finally, the scholarly impact of papers and authors related to retracted papers or authors seems not to be significantly affected, as long as those related papers and authors are not involved in any retraction cases.
\begin{table}
  \centering
  \scalebox{0.8}{
             \begin{tabular}{c|c|c|c} \hline
                Treatment group  	&No. of samples	&Post-retraction impact comparison		&Impact change ratio comparison\\
                \hline\hline
                $P^t$				&2659			&2 $<$ 6**						&0.5 $<$ 1.12**\\			
                $A^t$				&2092			&69.5 $<$ 80.5**					&0.76 $<$ 0.93**\\
                $I^t$				&2035			&670 $>$ 375**    					&3.72 $>$ 2.48**\\
                $P^t_{citing}$   		&1889			&5 $=$ 5							&1.33 $>$ 1.22\\
                $P^t_{coref}$		&2335			&5 $>$ 4*							&1 $=$ 1\\
                $A^t_{coaut}$		&1915			&285 $>$ 261*						&0.62 $>$ 0.60\\
                \hline
             \end{tabular}}\centering
      		\caption{Results of comparative tests. In the last two columns, the numbers before and after comparison symbols are median values from treatment groups and control groups, respectively. In addition, the mark * indicates p-value less than 0.05; the mark ** indicates p-value less than 0.01.}
        \label{tab:compare}
\end{table}

\subsubsection*{\textit{Segmentation analysis}}
Figure~\ref{fig:reason} shows the distribution of papers by reason for retraction. We also classify scholars and institutions by reasons for retraction based on their retracted papers. In this section, we look at the effect that the reason for retraction itself may have on the scholarly impact of papers and authors. We hypothesize that certain types of reasons for retraction may have more impact than others. 

In addition, we want to examine whether the increased attention brought by media coverage further decreases the scholarly impact of papers and authors involved in retractions. To do this, we extracted notable cases of scientific misconduct (most of them are falsification cases) reported by mass media and highlighted in Wikipedia\footnote{\url{https://en.wikipedia.org/wiki/Scientific_misconduct}}. These cases involved a total of 30 scholars and their retracted papers, all which can be found in our WoS corpus.

Table~\ref{tab:seg} compares the median values of the impact change ratio for retracted publications and authors, organized by retraction reason. The smaller the value, the greater the decrease in scholarly impact after retraction. We can see that retractions due to falsification or fabrication have the most substantial decrease in their scholarly impact for both papers and authors. This decrease is even more pronounced when the retraction cases are exposed to the public by media.
 

\begin{table}
  \centering
  \scalebox{0.7}{
             \begin{tabular}{c|c|c|c|c|c|c} \hline
                Retracted entity 	&Overall	&Misconduct-Media covered	&Misconduct-Falsification	 &Misconduct-Plagiarism	 &Misconduct-Violation	&Error\\
                \hline\hline
                $P^t$ 			&0.5		&\textbf{0.23}			&\textbf{0.33}				 &0.5					 &0.5					&0.5\\
                $A^t$				&0.76	&\textbf{0.31}			&\textbf{0.54}				 &0.77				 &0.58 				&0.79\\
               \hline
             \end{tabular}}\centering
      		\caption{Segmentation analysis of scholarly impact change ratio due to different retraction reasons. Retraction due to falsification or fabrication causes the most serious decrease of scholarly impact, and media coverage further exacerbates this trend.}
        \label{tab:seg}
\end{table}

\subsection*{\textit{Granger-causality analysis for effect of retraction on topical popularity }}
Ideally, we could apply the same comparative method to investigate the effect of retraction on scientific topics. However, it turns out to be very difficult to find a set of control topics that are comparable to the treatment topics of retracted papers. Instead, we apply the Granger causality test to examine whether the yearly topical retraction rate ($Ret^k(y)$) could statistically cause, or in other words, predict, the value of future yearly topical popularity ($Pop^k(y)$), for the top ten retracted topics listed in Table~\ref{tab:top}. 

The \emph{Granger Causality Test} is a statistical hypothesis test for determining whether one time series is useful in forecasting another. Specifically, a time series $X(t)$ is said to Granger-cause $Y(t)$ if it can be shown, usually through a series of t-tests and F-tests, previous $X(t)$ values provide statistically significant information about future values of $Y(t)$~\cite{tinfah:Granger1969Investigating}. As is shown in Equation~\ref{eq:granger}, the Granger Causality test can be mathematically formulated as a linear regression model, 
where $Y(t)$ is the predicted variable, $X(t)$ is the predicting variable, $C$ is a constant, $E(t)$ is the error term, $A_i$ and $B_j$ are corresponding coefficients, $n$ is the number of lagged years.
\begin{equation}
Y(t) = \sum^{n}_{i=1}A_iY(t-i) + \sum^{n}_{j=1}B_jX(t-j) + C + E(t)
\label{eq:granger}
\end{equation}
In our case, the Granger Causality test can help us to determine whether $Ret^k(y)$ Granger-causes a change of $Pop^k(y)$. In other words, whether retractions in a topic lead to an overall decrease in that topic's future popularity.

Table~\ref{tab:granger} lists the p-values of Granger Causality tests predicting $Pop^k(y)$ in the top ten most-frequently retracted topics, based on Equation~\ref{eq:granger} and set $n=1,2,3$ respectively. The rest of less frequently occurred retracted topics are ignored, to avoid the effect of data sparsity problem on the correctness of tests.
Most of the test results are not significant (i.e. p-value $>$ 0.05), except for three topics: \emph{gene expression}, \emph{apoptosis} and \emph{cell}. After checking the corresponding coefficients of previous popularity $Pop^k(y-i)$ (i.e. $A_i$), and previous retraction rate $Ret^k(y-j)$ (i.e. $B_j$) from Table~\ref{tab:cof}, we can see that the magnitude of $B_j$ is close to zero and much smaller than $A_i$. This suggests that the effect of historical retraction rate on the topical popularity is almost negligible, comparing with that of historical popularity, even if such a tiny effect seems to be statistically significant. 
Overall, the results from Granger Causality tests show that retraction rate in one topic hardly affects its future popularity. 
%
%
\begin{table}
  \centering
  \scalebox{0.8}{
             \begin{tabular}{c|c|c|c} \hline
                Topic 							&n=1						&n=2						&n=3\\
                \hline\hline
                regulation of gene expression 			&0.4							&0.29						&0.34\\
                gene expression					&0.57						&\textbf{0.01}					&\textbf{0.004}\\
                protein				 			&0.5							&0.3							&0.22\\
                apoptosis							&\textbf{0.01}					&0.16						&0.39\\
                cell (biology)						&\textbf{0.02}					&0.19						&0.12\\
                gene								&0.24						&0.17						&0.56\\
                enzyme inhibitor					&0.64						&0.98						&0.99\\
                cancer							&0.15						&0.45						&0.69\\
                tumor								&0.74						&0.66						&0.97\\
                inflammation						&0.1							&0.65						&0.36\\
               \hline
             \end{tabular}}\centering
      		\caption{Granger causality test of the possibility that retraction rate $Ret^k(y)$ affects topical popularity $Pop^k(y)$.}
        \label{tab:granger}
\end{table}

\begin{table}
  \centering
  \scalebox{0.58}{
             \begin{tabular}{c|c|c|c} \hline
                Topic 				&$n=1$	&$n=2$									&$n=3$\\
                \hline\hline
                gene expression		&NA		&$A_1=1.20,A_2=-0.22;B_1=-0.003, B_2=-0.005$, 	&$A_1=1.26,A_2=-0.01,A_3=-0.33;B_1=-0.006, B_2=-0.005, C_2=0.002$\\
                apoptosis				&$A_1=0.94;B_1=0.006$		&NA						&NA\\
                cell (biology)			&$A_1=0.91:B_1=-0.001$		&NA						&NA\\
               \hline
             \end{tabular}}\centering
      		\caption{Coefficients of Granger Causality Tests with significant p-values. $A_i$ represents the coefficient of $Pop^k(y-i)$ and $B_j$ represents the coefficient of $Ret^k(y-j)$}
        \label{tab:cof}
\end{table}

\section*{Discussion and Conclusion}
The ongoing evolution toward a global digital library and online publishing has opened a new era for scholarly communication. This has enabled easier access for reading and downloading scholarly publications by more people, exposing articles to public examination and investigation at an unprecedented scale and speed. At least in part, this increased readership may have resulted in more frequent paper retractions, and in increased awareness of retractions by both the scientific community and the popular media.

Some common themes related to retractions, as we raised at the start of the paper, are investigated and discussed: What does a typical retracted paper look like? How does the scientific community react to paper retractions? To what extent does an increase in retractions impact academia and science? To investigate the above questions, we conducted a comprehensive study and exploration of a set of retracted articles from Thomson Reuters Web of Science (WoS) across 30 years. Our main findings are summarized as follows. 

First, papers are retracted for multiple reasons with the most frequent reason being scientific misconduct. Based on 1,666 annotated retracted articles in our study, scientific misconduct accounted for more than 50\% of retractions. The second most common reason for retraction was accidental errors, which comprise around 24\% of retractions. Within all retractions due to scientific misconduct, plagiarism occurs most frequently, followed by fabrication and falsification. With regards to the research subject, the top three subjects of retracted publications are molecular biology \& genetics, immunology, and microbiology. 

Second, the scholarly impact of retracted papers and their authors significantly decreases after retraction. The results of our comparative experiments show that retracted papers receive significantly fewer citations after retraction than control articles published in the same journal on the same date. In addition, the reputation of scholars listed as first authors on retracted articles decreases significantly in the post-retraction period compared with other scholars having similar reputation growth patterns and research focus before retraction. In particular, those scholars who have falsified or fabricated data in their research received an even more severe decrease in their reputation, and media coverage further aggravates this penalty. Decreased scholarly recognition, based on citation count, of retracted papers and scholars, reflects the punitive reaction of the scientific community to falsified or erroneous scientific studies.  

Third, even when retracted papers and their authors are penalized by the scientific community with fewer citations, this negative effect was not shown to propagate to their sponsoring institution, or to other related but innocent papers and scholars. According to our comparative experimental results, the reputation of those institutions that sponsored the scholars who were accused of scientific misconduct did not seem to be tarnished at all. In addition, neither was a propogation effect observed for papers topically related to retracted articles based on citation relationship, nor for those scholars related to retracted authors based on co-authorship. Moreover, the Granger Causality analysis of dynamic retraction rate to the overall popularity of a certain scientific topic shows that the phenomenon of retraction bears no apparent effect on the popularity of a research topic. All these results strongly suggest that the penalty effect of paper retractions is quite localized to the authors who are mainly responsible for the retraction and does not extend to the wider scientific community. 

A fundamental, yet controversial, question that remains with regards to paper retraction is: As the number of retraction incidences keeps increasing, is it a good or bad signal for the development of science? Some scholars may claim that the drastic increase in retractions suggests the prevalence of scientific misconduct which disobeys the principle of doing science and may harm the authority and activity of scientific research. Others may claim that paper retraction is just a normal mechanism of self-examination and self-correction inherent to the scientific community, and that the increasing rate of retraction indicates the enhancement of that mechanism, which actually benefits scientific development in the long run. Even though we cannot give definite preference to either opinion, our study shows that the increasing retraction cases do not shake the ``shoulders of giants''. Only those papers and scholars that are directly involved are shown to be impacted negatively by retractions. In contrast, the sponsoring research institutions, other related but innocent papers and scholars, and research topics are not negatively impacted by retraction. Therefore, from our perspective, while the phenomenon of retraction is worth the attention of academia, its scope of negative influence should not be overestimated.

Scholarly impact is only one aspect that is affected by retraction; there may exist other issues that are also influenced by retraction to be explored in future work. The retracted study from~\citet{vaccine}, shows that the effects of paper retraction may go beyond the academia and extend to other aspects of society. Secondly, we find in this study that citing retracted papers does not seem to lead to any serious academic consequences. We cannot provide details as to why this is the case, but hypothesize that full-text citation analysis may help us better understand the contextual information of where and how retracted papers were cited. 
By further analyzing the text surrounding the citations to retracted articles, it might be possible to understand the author's attitude and intention in citing the retracted study.

\section*{Acknowledgement}
We thank our colleagues William Dowling who provided Web of Science dataset and tool for authors name disambiguation, as well as James Pringle for his insightful comments on this manuscript.

\bibliographystyle{model5-names}
\bibliography{refs}

\end{document}